\documentclass[aps,prc,preprint,showpacs,floatfix,superscriptaddress]{revtex4-1}  % for review and submission
\usepackage{graphicx}
\usepackage{amsmath}
\usepackage{multirow}

\newcommand {\nc} {\newcommand}
\nc {\beq} {\begin{eqnarray}} \nc {\eol} {\nonumber \\} \nc {\eeq}
{\end{eqnarray}} \nc {\eeqn} [1] {\label{#1} \end{eqnarray}} \nc
{\eoln} [1] {\label{#1} \\} \nc {\ve} [1] {\mbox{\boldmath $#1$}}
\nc {\rref} [1] {(\ref{#1})} \nc {\Eq} [1] {Eq.~(\ref{#1})} \nc
{\re} [1] {Ref.~\cite{#1}} \nc {\dem} {\mbox{$\frac{1}{2}$}} \nc
{\arrow} [2] {\mbox{$\mathop{\rightarrow}\limits_{#1 \rightarrow
#2}$}}
\begin{document}
\title{Study of the direct $^{16}{\rm O}(p, \gamma)^{17}{\rm
F}$ astrophysical capture reaction within a potential model
approach}
\author{E.M. Tursunov} \email{tursune@inp.uz} \affiliation {Institute of
Nuclear Physics, Academy of Sciences, 100214, Ulugbek, Tashkent,
Uzbekistan} \affiliation {National University of Uzbekistan, 100174
Tashkent, Uzbekistan}
\author{S.A. Turakulov}
\email{turakulov@inp.uz} \affiliation {Institute of Nuclear Physics,
Academy of Sciences, 100214, Ulugbek, Tashkent, Uzbekistan}

\begin{abstract}
A potential model is applied for the analysis of the astrophysical
direct nuclear capture process $^{16}$O(p,$\gamma)^{17}$F. The
phase-equivalent potentials of the Woods-Saxon form for the
p$-^{16}$O interaction are examined which reproduce the binding
energies and the empirical values of ANC for the $^{17}$F(5/2$^+$)
ground and $^{17}$F(1/2$^+$) ($E^*$=0.495 MeV) excited bound states
from different sources. The best description of the experimental
data for the astrophysical $S$ factor is obtained within the
potential model which yields the ANC values of 1.043 fm$^{-1/2}$ and
75.484 fm$^{-1/2}$ for the $^{17}$F($5/2^{+}$) ground and
$^{17}$F($1/2^{+}$) excited bound states, respectively. The
zero-energy astrophysical factor $S(0)=9.321$ KeV b is obtained by
using the asymptotic expansion method of D. Baye. The calculated
reaction rates within the region up to 10$^{10}$ K are in good
agreement with those from the R-matrix approach and the Bayesian
model in both absolute values and temperature dependence.
\end{abstract}


 \maketitle

\section{Introduction}

\par It is well-known that the $^{16}$O(p,$\gamma)^{17}$F reaction
is one of the important processes of the CNO  cycle in the Hot
bottom hydrogen burning stellar nucleosynthesis, specifically in
most massive Asymptotic Giant Branch (AGB)
stars~\cite{rolf1988,il2015}. Due to the low Q-value of the
$^{16}$O(p,$\gamma)^{17}$F reaction (Q = 0.6005 MeV) an equilibrium
is quickly established between the abundances of $^{17}$F and
$^{16}$O. This sequence is quite sensitive to the estimation of the
$^{17}$O/$^{16}$O abundances ratio in the A$<$20 mass
region~\cite{clay2003}. Therefore, a precise determination of the
reaction rate of this direct capture process is a primary goal of
nuclear astrophysics since it influences the evolution of the
stellar system~\cite{gagl99}.

\par The first direct experimental measurements of this reaction were performed at the end of
1950s \cite{hester58,tanner59}. Direct measurement of the cross
section of nuclear reactions at very low energies, relevant to the
temperature conditions inside stars, is difficult due to very strong
Coulomb repulsion of colliding charged particles. In spite of these
conditions, the cross-section of the proton capture reaction outside
a closed $^{16}$O core was obtained in direct way in Refs.~
\cite{rolfs73,chow75,morlock97,il12}. At the same time, in Ref.
~\cite{artem09} the low energy data was obtained by the indirect
method in experimentally inaccessible astrophysical energy region.
As to the theory, the astrophysical $^{16}$O(p,$\gamma)^{17}$F
direct capture reaction has been studied in the framework of several
models ~\cite{smith2023}, such as a phenomenological R-matrix
approach ~\cite{azuma10,il08,phuc21}, potential cluster model
~\cite{dub2018}, Gamow shell model ~\cite{benn10}, Bayesian
approximation ~\cite{il22}, microscopic cluster model
~\cite{baye98}, and self-consistent mean-field potential model based
on the Hartree-Fock method \cite{nguen21}. One of the last
theoretical research works on this reaction has been performed
within the Bayesian fitting method \cite{il22} on the basis of the
single particle potential model taking into account statistical and
systematic uncertainties.

In Ref. ~\cite{gagl99} the empirical values of the asymptotic
normalization coefficient (ANC) extracted from the analysis of the
$^{16}\rm{O}(^3\rm{He,d})^{17}$F proton transfer reaction within the
distorted-wave Born approximation (DWBA) have been used for the
estimation of the astrophysical $S$ factor of the
$^{16}$O(p,$\gamma)^{17}$F direct capture reaction. Furthermore,
there are several works ~\cite{blokh18,barbieri05} devoted to the
extraction of the empirical value of ANC for the virtual decay
$^{17}$F $\to$ $^{16}$O+p in the $^{17}$F(5/2$^+$) ground and
$^{17}$F(1/2$^+$) excited bound states from different proton
transfer reactions. The empirical squared ANC values of
C$^{2}_{5/2^+}=1.028\pm$ 0.131 fm$^{-1}$ and
C$^{2}_{1/2^+}=5430\pm$950 fm$^{-1}$ were extracted within the frame
of the modified DWBA~\cite{feruz22} by analyzing the
$^{16}\rm{O}(^{10}\rm{B},^9\rm{Be})^{17}$F reaction at the energy
E$_{^{10}\rm{B}}$ = 41.3 MeV. Other ANC estimations were performed
by the analysis of the peripheral nuclear reaction
$^{16}\rm{O}(^3\rm{He,d})^{17}$F  at different energies of $^3$He
ions in DWBA ~\cite{gagl99,artem09}. The extracted values of ANC for
the ground state $^{17}$F(5/2$^+$) are practically the same,
C$^{2}_{5/2^+}=1.08\pm$ 0.10 fm$^{-1}$ ~\cite{gagl99} and
C$^{2}_{5/2^+}=1.09\pm$ 0.11 fm$^{-1}$ ~\cite{artem09},
respectively. On the other hand, for the first excited bound state
$^{17}$F(1/2$^+$) (E$^*$=0.495 MeV) the squared ANC value
C$^{2}_{1/2^+}=6490\pm$680 fm$^{-1}$ of Ref. ~\cite{gagl99} differs
significantly from the estimate C$^{2}_{1/2^+}=5700\pm$225 fm$^{-1}$
reported in Ref.~\cite{artem09}.

The aim of the present work is to perform a comparative analysis of
different potential models for the description of the
$^{16}$O(p,$\gamma)^{17}$F direct capture reaction. Our goal is to
find the parameters of the successful realistic potential model
which can describe the existing experimental data for the
astrophysical $S$ factor at low energies
\cite{rolfs73,chow75,morlock97,artem09,il12}. These potential models
should describe the both bound-state properties of light nuclei
(binding energies, ANC), as well as scattering data (experimental
phase shifts, scattering lengths) \cite{mukh2016,gnec2019}.

Specifically, proposed potential models of the Woods-Saxon form will
be able to reproduce the empirical ANC values for the
$^{17}$F(5/2$^+$) ground and $^{17}$F(1/2$^+$) excited bound states
deduced in Refs.~\cite{feruz22,artem09,gagl99,il22} in addition to
the experimental bound state energies. The potential parameters in
the $^2D_{5/2}$ and $^2S_{1/2}$ scattering waves, if necessary, are
adjusted to reproduce the experimental phase-shifts
\cite{blue65,chow75}. Then the potential models will be examined in
terms of their ability to describe the experimental astrophysical
$S$ factor and the reaction rates of the $^{16}$O(p,$\gamma)^{17}$F
direct capture process.

The theoretical model will be briefly described in Section II, the
numerical results are presented in Section III and the conclusions
are given in the last section.

\section{Theoretical model}

\subsection{Wave functions}

The wave functions of the initial scattering and final bound states
within a single-channel potential model
~\cite{tur18,tur2021a,tur2021b,tur2023a} are presented as
\begin{eqnarray}
\Psi_{lS}^{J}=\frac{u_E^{(lSJ)}(r)}{r}\left\{Y_{l}(\hat{r})\otimes\chi_{S}(\xi)
\right\}_{J M}
\end{eqnarray}
and
\begin{eqnarray}
\Psi_{l_f S'}^{J_f}
=\frac{u^{(l_fS'J_f)}(r)}{r}\left\{Y_{l_f}(\hat{r})\otimes\chi_{S'}(\xi)
\right\}_{J_f M_f},
\end{eqnarray}
respectively.

The solutions of the two-body Schr\"{o}dinger equation
\begin{align}
\left[-\frac{\hbar^2}{2\mu}\left(\frac{d^2}{dr^2}-\frac{l(l+1)}{r^2}\right)+V^{
lSJ}(r)\right] u_E^{(lSJ)}(r)= E u_E^{(lSJ)}(r),
\end{align}
represent the radial wave functions in the initial $^2S_{1/2}$,
$^2P_{1/2}$, $^2P_{3/2}$, $^2D_{5/2}$ scattering states of the
p-$^{16}${\rm O} system, where $V^{lSJ}(r)$ is a two-body potential
in the partial wave with orbital angular momentum $l$, spin $S$ and
total angular momentum $J$. The wave function $u^{(l_fS'J_f)}(r)$ of
the final $^2D_{5/2}$ ground and $^2S_{1/2}$ excited states are
solutions of the bound-state Schr\"{o}dinger equation.

The p-$^{16}${\rm O} two-body potential, containing the central and
the spin-orbit terms, is chosen in the Woods-Saxon form
\cite{huang10}:
\begin{eqnarray}
V^{lSJ}(r)=V_{0}\left[1+exp\left(\frac{r-R_0}{a_0}\right)\right]^{-1}-\\
\nonumber -V_{SO}\left(\frac{\hbar}{m_{\pi}c}\right)^2(\vec{l}\cdot
\vec{S})\frac{1}{r}\frac{d}{dr}\left[1+exp\left(\frac{r-R_{SO}}{a_{SO}}\right)\right]^{-1}+V_{c}(r),
\label{pot}
 \end{eqnarray}
where $V_0$ and $V_{SO}$ are depths of the central and spin-orbit
parts of the potential, $R_0=R_{SO}=r_0 \,A^{1/3}$ fm and
$a_0=a_{SO}$ are geometric parameters of the potential radius and
diffuseness, respectively. Here, $A_1$ and $A_2$ are the atomic mass
numbers of the first (proton) and second ($^{16}${\rm O} nucleus)
clusters, respectively, and $A=A_1+A_2$. The Coulomb potential is
given by the spherical charge distribution \cite{huang10}
\begin{eqnarray}
 V_c(r)=
\left\{
\begin{array}{lc}
 Z_1 Z_2 e^2/r &  {\rm if} \,\, r>R_c, \\
Z_1 Z_2 e^2 \left(3-{r^2}/{R_c^2}\right)/(2R_c) & {\rm if}
\,\,r<R_c,
\end{array}
\right. \label{Coulomb}
\end{eqnarray}
with the Coulomb radius $R_c=1.25 A^{1/3}$ fm, and charge numbers
$Z_1$, $Z_2$ of the first and second clusters, respectively.

\subsection{Cross sections of the radiative-capture process}

In the present paper, the total cross-section of the radiative
capture process is represented by the sum of the cross-sections for
all the final states \cite{tur2021b,tur2023a}:
\begin{eqnarray}
\sigma(E)=\sum_{J_f \lambda \Omega}\sigma_{J_f \lambda}(\Omega),
\end{eqnarray}
where $\Omega=$ $E$ (electric) or $M$ (magnetic) transitions,
$\lambda$ is the transition multiplicity, $J_f$ is the total angular
momentum of the final state. For a particular final state with total
angular momentum $J_f$ and multiplicity $\lambda$ we have
\cite{tur2021a}
\begin{align}
 \sigma_{J_f \lambda}(\Omega) =& \sum_{J}\frac{(2J_f+1)} {\left
[S_1 \right]\left[S_2\right]} \frac{32 \pi^2 (\lambda+1)}{\hbar
\lambda \left( \left[ \lambda \right]!! \right)^2} k_{\gamma}^{2
\lambda+1} C^2(S) \sum_{l S}
 \frac{1}{ k_i^2 v_i}\mid
 \langle \Psi_{l_f S'}^{J_f}
\|M_\lambda^\Omega\| \Psi_{l S}^{J} \rangle \mid^2,
\end{align}
where $l$ and $l_{f}$ are the orbital momenta of the relative motion
in the initial and final states, respectively; $k_i$ and $v_i$ are
the wave number and speed of the p-$^{16}${\rm O} relative motion in
the entrance channel; $S_1$ and $S_2$ are spins of $p$ and
$^{16}${\rm O} cluster,  $S'=S=$1/2 due to the use of the
single-channel approximation. The wave number of the photon
$k_{\gamma}=E_\gamma / \hbar c$ corresponds to the energy
$E_\gamma=E_{\rm th}+E$, where $E_{\rm th}$ is the threshold energy
for the breakup reaction $\gamma+^{17}$F$\to^{16}$O+p. Constant
$C^2(S)$ is a spectroscopic factor \cite{NACRE99}, which is equal to
1 in the potential model according to Ref. \cite{mukh2016} .

The reduced matrix elements of the transition operators are
calculated between the initial $\Psi_{l S}^{J}$ and final $\Psi_{l_f
S'}^{J_f}$ states. In the long-wavelength approximation the electric
transition operator can be written in the form
\begin{eqnarray}
M_{\lambda \mu}^{\rm E}=e \sum_{j=1}^{A}
Z_j{r'_j}^{\lambda}Y_{\lambda \mu}(\hat{r'}_j),
\end{eqnarray}
where $\vec {r'}_{j}= \vec{r}_j-\vec{R}_{cm}$ is the position of the
$j$-th particle in the center of mass system.

The analytical expression for the reduced matrix elements of this
operator can be written as \cite{tur2021a,tur2023a}
\begin{eqnarray}
\langle \Psi_{l_f S'}^{J_f}\|M_\lambda^{\rm E}\| \Psi_{l S}^{J}
\rangle
 &=& e\left[Z_1 \left( \frac{A_2}{A} \right)^{\lambda}+Z_2
\left(\frac{-A_1}{A} \right)^{\lambda} \right]   \delta_{S S'} \\
\nonumber && \times (-1)^{J+l+S}\left(\frac{[\lambda][l][J]}{4
\pi}\right)^{1/2} C^{l_f 0}_{\lambda 0 l 0} \left\{
\begin{array}{ccc}
J & l & S \\
l_{f} & J_{f} & \lambda
\end{array} \right\} \\ \nonumber && \times  \int^{\infty}_{0} u_{E}^{(lSJ)}(r)r^{\lambda} u^{(l_fSJ_f)} (r) dr,
\end{eqnarray}
where $A_1$, $A_2$  are the mass numbers of the clusters in the
entrance channel, $A=A_1+A_2$.

The magnetic transition operator can be written as the sum of the
orbital and spin parts \cite{tur18,tur2021a}:
\begin{eqnarray}
M_{1 \mu}^M&=& \sqrt{\frac {3}{4 \pi}}\left[ \sum_{j=1}^{A}
 \mu_N \frac{Z_j}{A_j}\hat{l}_{j \mu} + 2 \mu_j \hat{S}_{j \mu} \right]
 \nonumber \\
 &=& \sqrt{\frac {3}{4 \pi}} \left[\mu_N \left( \frac{A_2 Z_1}{A A_1} + \frac{A_1
 Z_2}{A A_2} \right) \hat{l}_{r \mu} +2(\mu_1\hat{S}_{1\mu}+
 \mu_2\hat{S}_{2\mu})\right],
\end{eqnarray}
where $\mu_N$ is the nuclear magneton, $\mu_j$ is the magnetic
moment, $\hat{l}_{j \mu}$ is the orbital momentum of the $j-$th
particle and $\hat{l}_{r \mu}$ is the orbital momentum operator of
the relative motion of the two clusters.
%
%It is established according to previous studies for the two-cluster
%system in the case of spins of cluster nonequal zero \cite{tur2021a}
%and in the case of one cluster spin equal zero \cite{tur18}.
The reduced matrix elements of the magnetic M1 transition operator
can be evaluated using the values of the cluster spins
$S_1=S_p=1/2=S$ and $S_2=S(^{16} O)$=0 \cite{tur18,tur2021a}   
\begin{eqnarray}
\langle \Psi_{l_f S}^{J_f}\|M_1^M\| \Psi_{l S}^{J} \rangle & = &
\mu_N \left( \frac{A_2 Z_1}{A A_1} + \frac{A_1
 Z_2}{A A_2} \right) \sqrt{l_f(l_f+1)[J_f][l_f]}(-1)^{S+1+J_f+l_f}
\left\{
\begin{array}{ccc}
l_f & S & J_f \\
J & 1 & l_f
\end{array}
\right\} I_{if}  \nonumber   \\
&&+  2 \mu({\rm p})(-1)^{1+l_f+S+J} \sqrt{S(S+1)[S][J_f]} \left\{
\begin{array}{ccc}
S & l_f & J_f \\
J & 1 & S
\end{array}
\right\}  I_{if},
\end{eqnarray}
where magnetic moment of the proton $\mu_{p}=$2.792847$\mu_N$  and
the overlap integral is given as
\begin{eqnarray}
I_{if}= \delta_{l l_f} \sqrt{\frac {3}{4 \pi}} \int^{\infty}_{0}
u_{E}^{(lSJ)}(r)u^{(l_fSJ_f)} (r) dr.
\end{eqnarray}

Finally, the cross section and the astrophysical $S$ factor of the
process are related to each other by the equation \cite{fow1975}
\begin{eqnarray}
S(E)=E \, \, \sigma(E) \exp(2 \pi \eta).
\end{eqnarray}

\section {Numerical results}

\subsection{Astrophysical $S$ factor of the $^{16}$O(p,$\gamma)^{17}$F direct capture reaction}

The Schr\"{o}dinger equation is solved numerically in the entrance
and exit channels with the two-body $p-^{16}\rm{O}$ nuclear
potentials of the Woods-Saxon form (\ref{pot}) with the
corresponding Coulomb part of the spherical charge distribution.
Hereafter, everywhere the parameter values within the atomic mass
units $\hbar^2/2 u$ =20.9008 MeV fm$^2$, 1u=931.494 MeV, $m_p
=A_1$u=1.007276466812u , m($^{16}$O)=$A_2$u=15.994915u and
$\hbar$c=197.327 MeV fm are used in numerical calculations.

\begin{table}[htbp]
\centering \caption{Potential parameters for the p-$^{16}$O
interaction in different potential models.} \label{table1}
\begin{tabular}{c c c c c c c c}
\hline & $^{2S+1}L_J$ & $V_{0}$ (MeV)  & $V_{SO}$ (MeV) & $r_0$ (fm) & $a_0$ (fm) & $C_{LJ}$ (fm$^{-1/2})$ &$E_{FS}^{^{17}\rm{F}}$ (MeV)\\
\hline
\multirow{2}{3em}{$\textrm{V}_\textrm{M1}$} & $^2S_{1/2}$ & -58.7666 & 2.8 & 1.17 & 0.552 & 73.404\cite{feruz22} & -31.65 \\
& $^2D_{5/2}$ & -55.2722 & 4.6 & 1.23 & 0.699 & 1.012 \cite{feruz22} &- \\
\hline
\multirow{2}{3em}{$\textrm{V}_\textrm{M2}$} & $^2S_{1/2}$ & -57.3484 & 2.8 & 1.18 & 0.585 & 75.484\cite{artem09} &-30.32 \\
& $^2D_{5/2}$ & -54.4731 & 4.6 & 1.24 & 0.716 & 1.043\cite{artem09} &- \\
\hline
\multirow{2}{3em}{$\textrm{V}_\textrm{M3}$} & $^2S_{1/2}$ & -53.8711 & 2.8 & 1.21 & 0.656 & 80.450\cite{gagl99} &-27.42 \\
& $^2D_{5/2}$ & -54.4780 & 4.6 & 1.24 & 0.712& 1.038\cite{gagl99} &- \\
\hline
\multirow{2}{3em}{$\textrm{V}_\textrm{M4}$} & $^2S_{1/2}$ & -52.2137 & 2.8 & 1.22 & 0.712 & 84.025\cite{il22} &-25.73 \\
& $^2D_{5/2}$ & -54.4570 & 4.6 & 1.24 & 0.727 & 1.056\cite{il22} &-
\\\hline
\end{tabular}
\vspace*{0.5cm}
\end{table}

The parameters of the central and spin-orbital parts of  Woods-Saxon
potential for different models are presented in Table \ref{table1}.
The supposed potential models
$\textrm{V}_\textrm{M${\rm{1}}$}$,$\textrm{V}_\textrm{M${\rm{2}}$}$,$\textrm{V}_\textrm{M${\rm{3}}$}$
and $\textrm{V}_\textrm{M${\rm{4}}$}$ differ from each other only in
the $^2D_{5/2}$ and $^2S_{1/2}$ partial waves which correspond to
the ground and first excited bound states of the $^{17}$F nucleus,
respectively. The depths of the potential in the $^2P_{1/2}$ and
$^2P_{3/2}$ initial scattering waves are taken equal to zero, which
means that the pure Coulomb scattering is considered due-to the
closed shell of the $^{16}$O cluster.

\begin{figure}[htbp]
\includegraphics[width=10 cm]{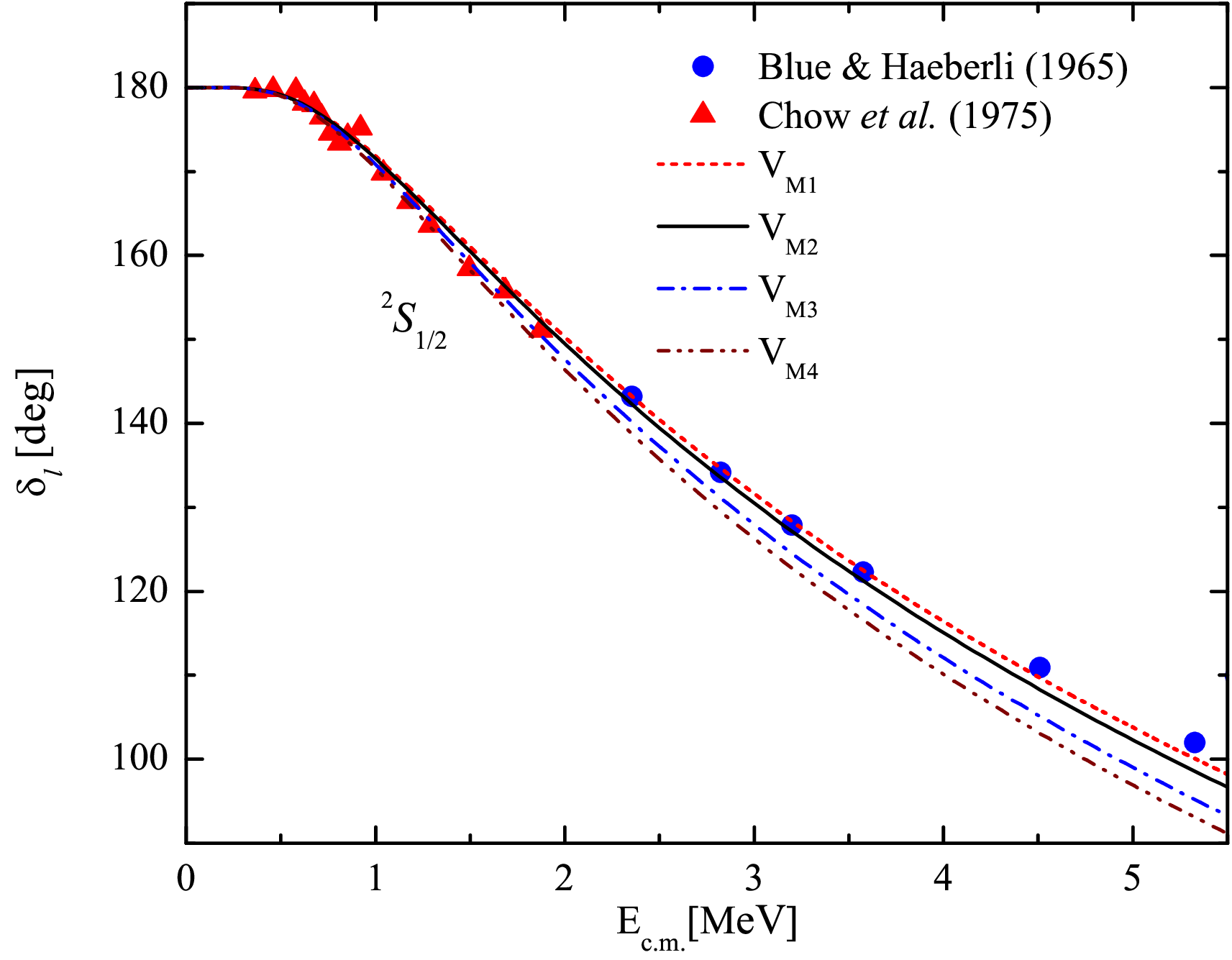} \caption{The $^2S_{1/2}$ wave nuclear
phase shifts for $p-^{16}${\rm O} elastic scattering within
different potential models. The experimental data are taken from
Ref.~\cite{blue65,chow75}.} \label{fig1}
\end{figure}

\begin{figure}[htbp]
\includegraphics[width=10 cm]{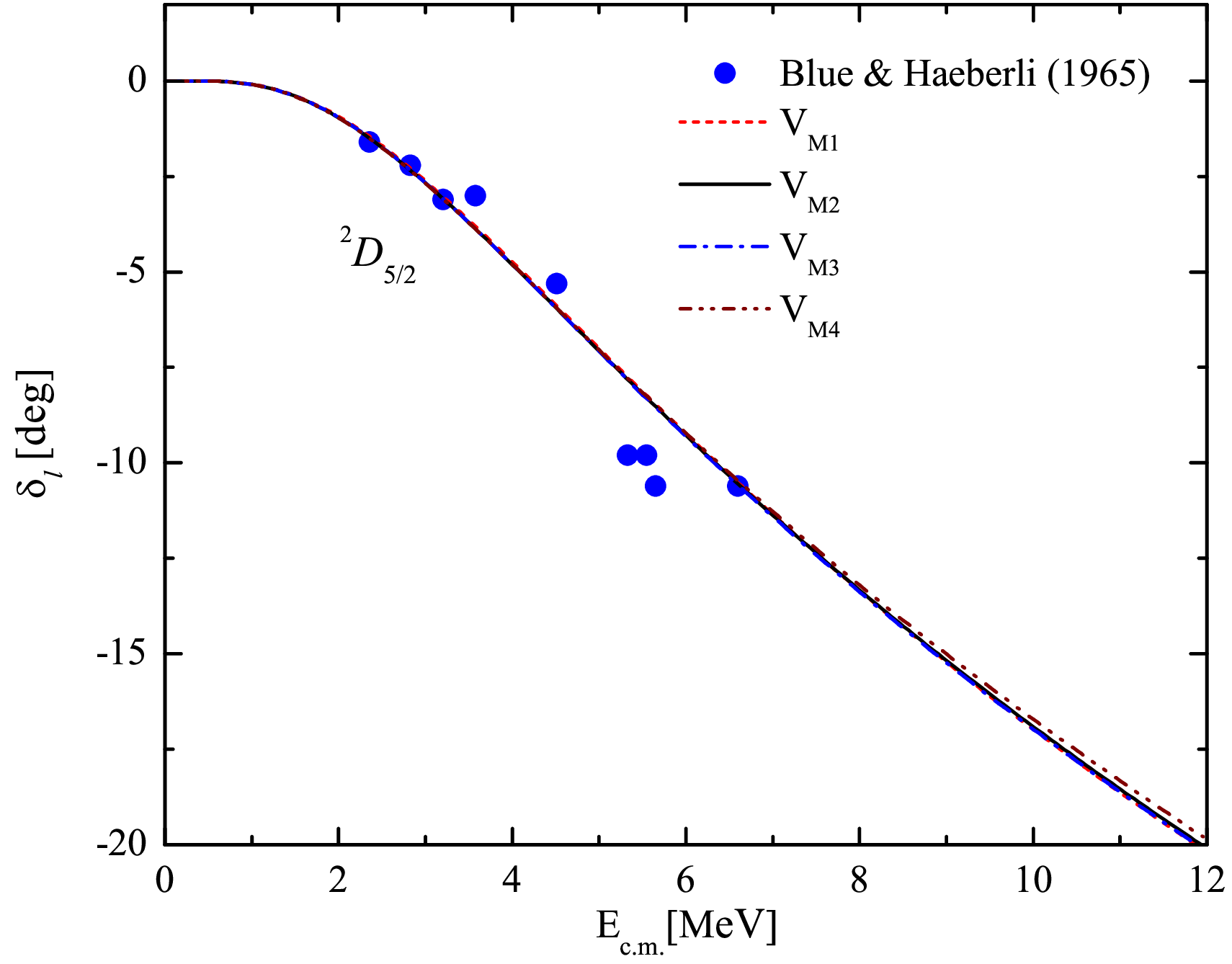} \caption{The $^2D_{5/2}$ wave nuclear
phase shifts for $p-^{16}${\rm O} elastic scattering within
different potential models. The experimental data are taken from
Ref.~\cite{blue65}.} \label{fig2}
\end{figure}

The parameters of the supposed $\textrm{V}_\textrm{M${\rm{1}}$}$,
$\textrm{V}_\textrm{M${\rm{2}}$}$,
$\textrm{V}_\textrm{M${\rm{3}}$}$, and
$\textrm{V}_\textrm{M${\rm{4}}$}$ potentials in the $^2D_{5/2}$ and
$^2S_{1/2}$ partial waves were fitted to reproduce the experimental
energies $E(5/2^+)$=-0.6005 MeV and $E(1/2^+)$=-0.1052 MeV and the
empirical ANC values from Refs. \cite{feruz22,artem09,gagl99,il22}
for the $^{17}$F(5/2$^+$) ground and $^{17}$F(1/2$^+$) excited bound
states, respectively. The presented values of the empirical ANC
values were obtained from the analysis of different experimental
data for the cross-section of the proton transfer reactions using
the modified DWBA approach. The same potential models describe the
experimental phase shifts quite well. In Fig.~\ref{fig1} the
calculated phase shifts are presented for the p$-^{16}$O scattering
in the $^2S_{1/2}$ partial wave using different sets of potential
models in comparison with the experimental data from Refs.
\cite{blue65,chow75}. Similar calculations have been performed for
the $^2D_{5/2}$ partial wave and obtained results are presented in
Fig.~\ref{fig2} in comparison with the available experimental data
from Ref. \cite{blue65}.

In the last column of Table \ref{table1} values of the forbidden
state energy in the $^2S_{1/2}$ partial wave for different potential
models are presented. For the $p-^{16}$O relative motion, the Pauli
forbidden state in the $S$-wave corresponds to the orbital
configuration [$s^5p^{12}$] of the constituent nucleons
~\cite{baye98}.

\begin{figure}[htbp]
\includegraphics[width=10 cm]{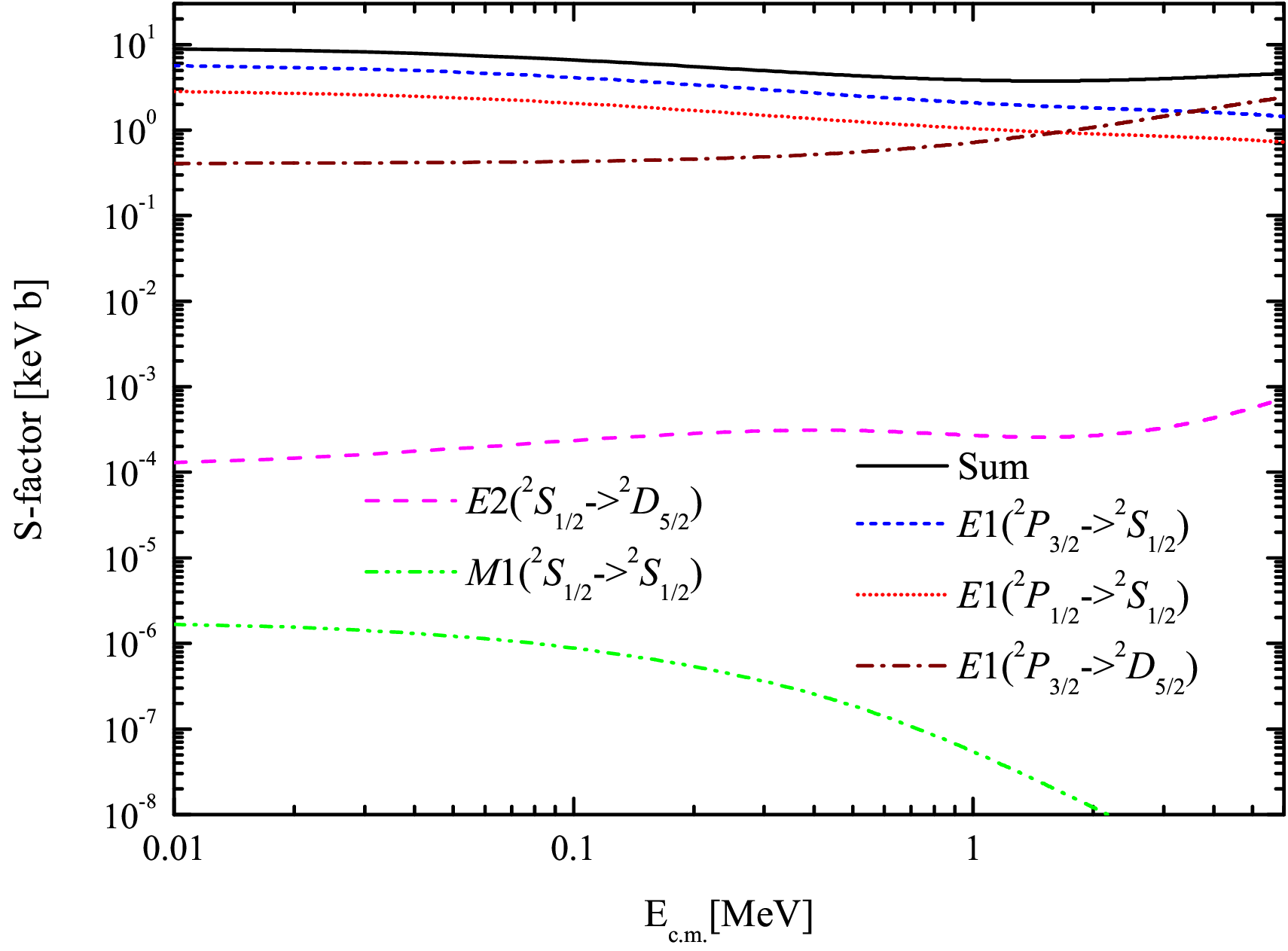}\caption{The $E$1, $E$2 and $M1$ partial
transition contributions to the astrophysical $S$ factor from
different initial scattering states to the final
$^{17}\rm{F(5/2^+)}$ ground and the first excited
$^{17}\rm{F(1/2^+)}$ bound states within the potential model
$\textrm{V}_\textrm{M${\rm{2}}$}$.} \label{fig3}
\end{figure}

In Figure~\ref{fig3} a comparison of the contributions of the
electromagnetic dipole and quadrupole transition operators into the
astrophysical $S$ factor of the $^{16}$O(p,$\gamma)^{17}$F synthesis
process is shown for the model $\textrm{V}_\textrm{M${\rm{2}}$}$
from Table \ref{table1}. As can be seen from the figure, the
electric $E$1 dipole transition yields a dominant contribution in
the energy region up to 6.0 MeV. It can be seen that for the $E$1
transitions the most important contribution comes from the initial
$P$-waves to the final $S$ state. The contributions of the electric
quadrupole ($E$2) and magnetic ($M$1) dipole transitions are much
more suppressed and differ from the contribution of the $E$1
transition by more than five and seven orders of magnitude in the
astrophysical low energy region up to $E$=1 MeV, respectively.

\begin{figure}[htbp]
\includegraphics[width=10 cm]{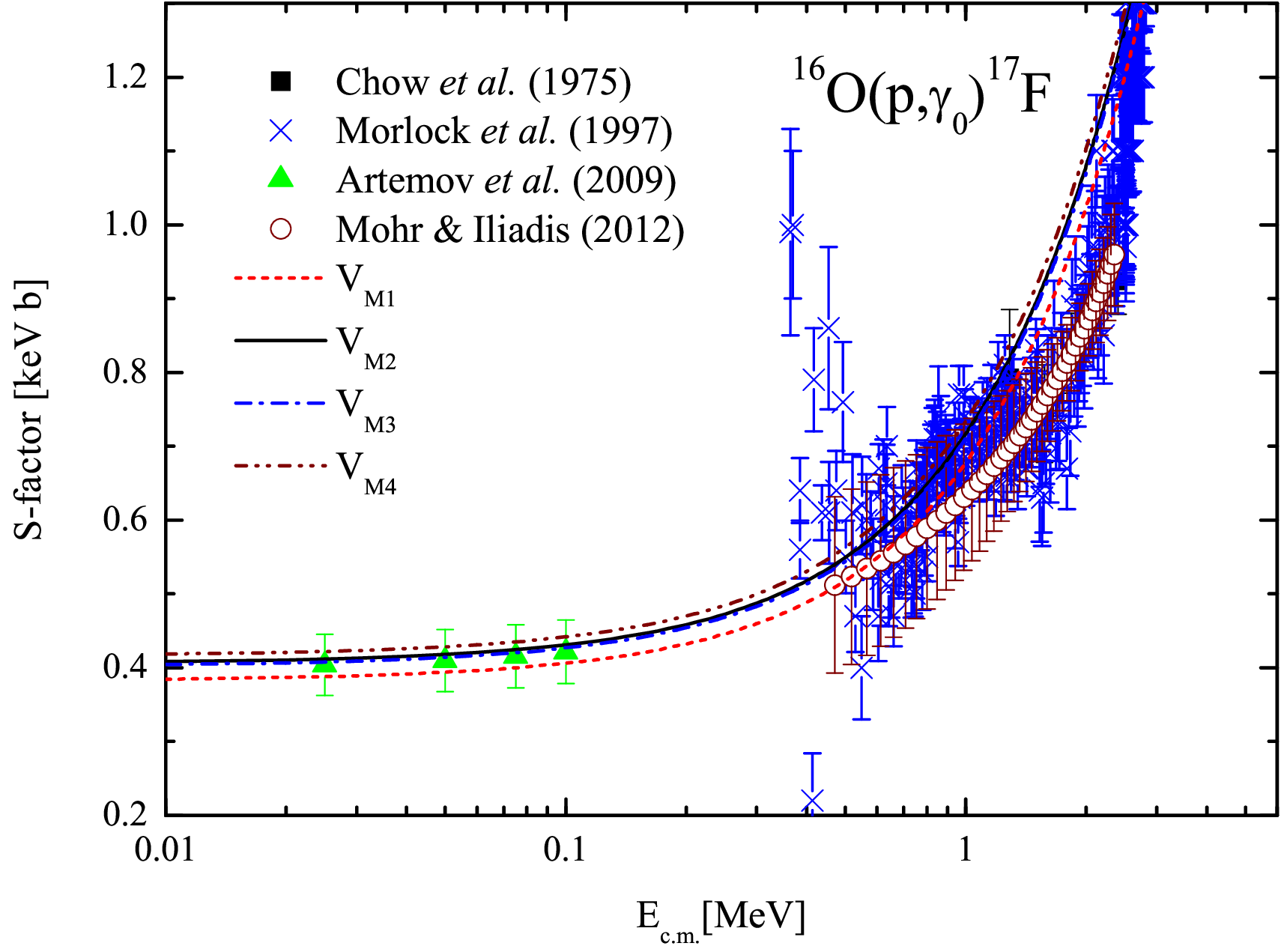}\caption{The $^{16}$O(p,$\gamma_0)^{17}$F astrophysical $S$
factors for the transition to the $^{17}\rm{F(5/2^+)}$ ground state,
calculated with different sets of potential models. The data are
taken from Refs.~\cite{artem09,chow75,morlock97,il12}. }
\label{fig4}
\end{figure}

In Fig.~\ref{fig4} the estimated astrophysical $S$ factors of the
direct radiative capture reaction
$^{16}$O(p,$\gamma_0)^{17}$F$(5/2^+)$ to the ground state with
different potential models are presented in comparison with the
experimental data sets from Refs.~
\cite{artem09,chow75,morlock97,il12}. As can be seen from
Fig.~\ref{fig4}, the astrophysical $S$ factors calculated within all
the potential models $\textrm{V}_\textrm{M${\rm{1}}$}$,
$\textrm{V}_\textrm{M${\rm{2}}$}$, $\textrm{V}_\textrm{M${\rm{3}}$}$
and $\textrm{V}_\textrm{M${\rm{4}}$}$ are in good agreement with the
available experimental data within the whole astrophysical
low-energy region. The most important result is the description of
the data from Ref.\cite{artem09} below 100 keV energy.

\begin{figure}[htbp]
\includegraphics[width=10 cm]{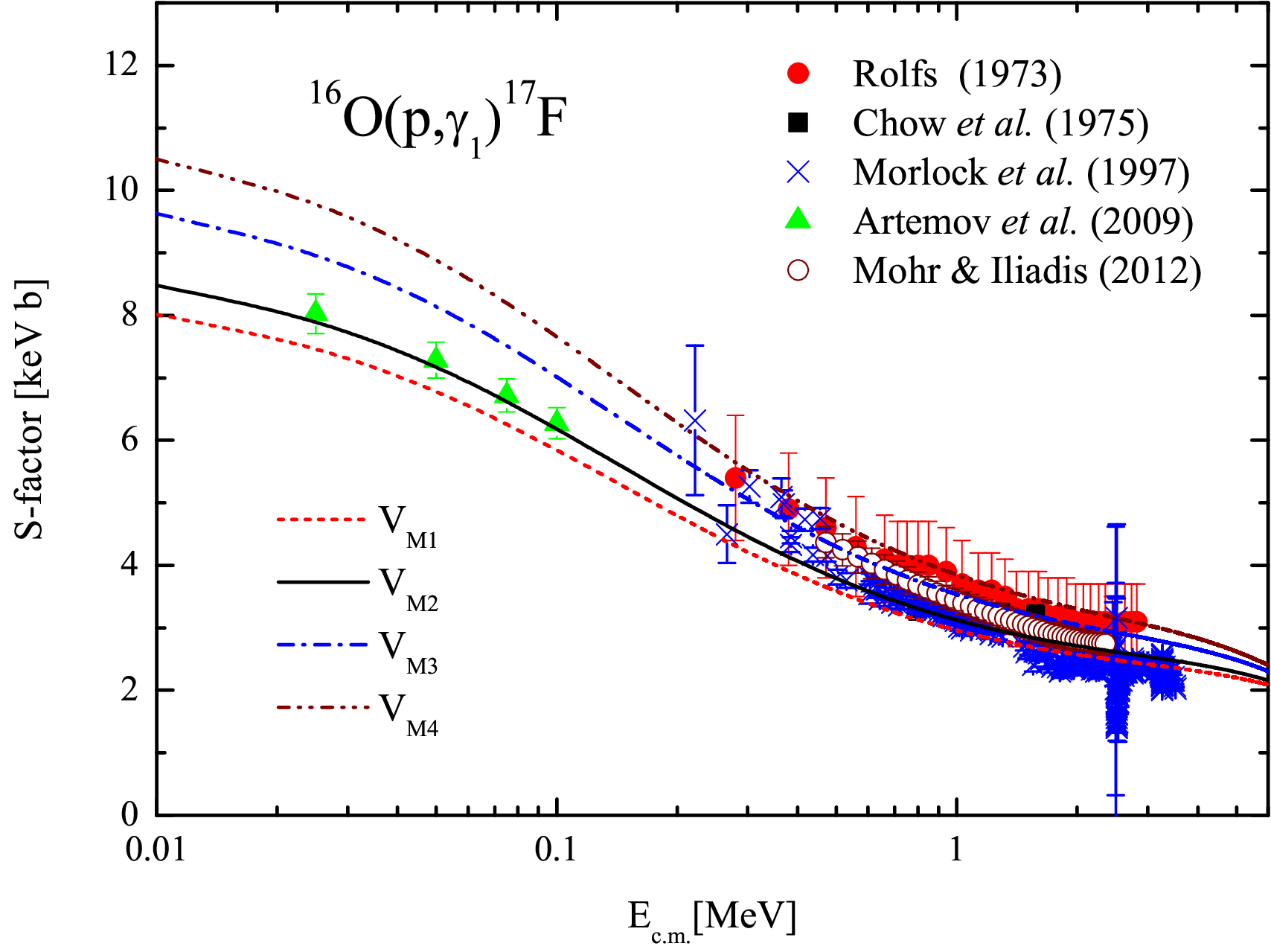}\caption{The $^{16}$O(p,$\gamma_1)^{17}$F astrophysical $S$
factors for the transition to the first excited $^{17}\rm{F(1/2^+)}$
bound state, calculated with different sets of potential models. The
data are taken from
Refs.~\cite{artem09,rolfs73,chow75,morlock97,il12}.} \label{fig5}
\end{figure}

In Fig.~\ref{fig5} similar comparative analysis is performed for the
transition to the first excited $^{17}\rm{F(1/2^+)}$ bound state
within the same potential models $\textrm{V}_\textrm{M${\rm{1}}$}$,
$\textrm{V}_\textrm{M${\rm{2}}$}$, $\textrm{V}_\textrm{M${\rm{3}}$}$
and $\textrm{V}_\textrm{M${\rm{4}}$}$, and the results are compared
with the experimental data sets from Refs.~
\cite{artem09,rolfs73,chow75,morlock97,il12}. In this case the best
description of the experimental data for the astrophysical $S$
factor of the $^{16}$O(p,$\gamma_1)^{17}$F$(1/2^+)$ direct capture
process corresponds to the $\textrm{V}_\textrm{M2}$ potential model,
which yields the ANC value of 75.484$\pm$1.490 fm$^{-1/2}$ for the
$^{17}\rm{F(1/2^+)}$ first excited bound state \cite{artem09}.

%$p-^{16}${\rm O} interaction in the
%for the $^{17}\rm{F(5/2^+)}$ ground and
%the first excited $^{17}\rm{F(1/2^+)}$ bound states.

In Table ~\ref{table2} we present the calculated values of  the
astrophysical $S(0)$ factor of the direct $^{16}$O(p,$\gamma)^{17}$F
capture process at the zero energy separately for the ground
$^{17}$F($5/2^{+}$) and first excited $^{17}$F($1/2^{+}$) bound
states and their sum. The zero-energy astrophysical $S(0)$ factor
was estimated by using the asymptotic expansion method
\cite{baye00,tur23b}. The result of the
$\textrm{V}_\textrm{M${\rm{2}}$}$ model is consistent with the
result of the ANC method of Ref.\cite{artem09} $S(0)=9.45 \pm 0.40$
keV b and with the NACRE compilation data $S(0)=9.3 \pm 2.8$ keV b
\cite{NACRE99}.

\begin{table}[htbp]
\centering \caption{The calculated values of the astrophysical
$S(0)$ factor at zero energy for the ground $^{17}$F ($5/2^{+}$) and
the first excited $^{17}$F($1/2^{+}$) ($E^*=0.495$ MeV) bound states
and their sum.} \label{table2}
 \begin{tabular}{c c c c c}
\hline Model & $\rm{V}_{\rm{M1}}$   & $\rm{V}_{\rm{M2}}$ & $\rm{V}_{\rm{M3}}$& $\rm{V}_{\rm{M4}}$ \\
E$^{\rm{*}}~\rm{(MeV)}$ & \multicolumn{4}{c}{ S(0)~\rm{(keV b)}}\\
\hline
0.0 & 0.382 & 0.406 & 0.402 &0.416\\
0.495 & 8.427 & 8.915& 10.119&11.045\\
Total & 8.809 & 9.321& 10.521&11.461\\
\hline
\end{tabular} \label{tab1a}
\vspace*{0.5cm}
\end{table}

\subsection{Reaction rates of the $^{16}$O(p,$\gamma)^{17}$F
astrophysical direct capture process}

\par A realistic estimate of the reaction rate is one of the important input
quantities for the calculation of abundances of chemical elements in
the Universe in the stellar evolution of the Big Bang model. The
reaction rate $N_{A}(\sigma v)$ is estimated with the help of the
expression \cite{NACRE99,fow1984}
\begin{eqnarray}
N_{A}(\sigma v)=N_{A}
\frac{(8/\pi)^{1/2}}{\mu^{1/2}(k_{\text{B}}T)^{3/2}}
\int^{\infty}_{0} \sigma(E) E \exp(-E/k_{\text{B}}T) d E,
\end{eqnarray}
where $\sigma(E)$ is the calculated cross-section of the process,
 $k_{\text{B}}$ is the Boltzmann coefficient, $T$ is the
temperature, $N_{A}=6.0221\times10^{23}\, \text{mol}^{-1}$ is the
Avogadro number.

\begin{table}[htbp]
\centering \caption{The $^{16}$O(p,$\gamma)^{17}$F astrophysical
reaction rates $N_{A}(\sigma v)$ ($ \textrm{cm}^{3} \rm{mol}^{-1}
\rm{s}^{-1}$) in the temperature interval $ 0.001\leq T_{9} \leq 10
$ calculated within the potential model $\textrm{V}_\textrm{M2}$.}
\bigskip
\begin{tabular}{c c c c c c c c} \hline
~~~$T_{{\rm{9}}}$ ~~~~& ~~ $\textrm{V}_\textrm{M2}$~~~~~~~&~~~
$T_{{\rm{9}}}$~~~ &~~~~$\textrm{V}_\textrm{M2}$~~~~\\
\hline
0.001&$4.66\times 10^{-63}$ &0.14&$4.02\times 10^{-6}$\\
0.002&$2.61\times 10^{-48}$ &0.15&$7.87\times 10^{-6}$\\
0.003&$3.70\times 10^{-41}$ &0.16& $1.46\times 10^{-5}$\\
0.004&$1.19\times 10^{-36}$ &0.18&$4.31\times 10^{-5}$ \\
0.005&$1.91\times 10^{-33}$ &0.20& $1.09\times 10^{-4}$\\
0.006&$5.32\times 10^{-31}$ &0.25&$7.01\times 10^{-4}$\\
0.007&$4.74\times 10^{-29}$ &0.30&$2.87\times 10^{-3}$\\
0.008&$1.92\times 10^{-27}$ &0.35&$8.77\times 10^{-3}$\\
0.009&$4.39\times 10^{-26}$ &0.40&$2.20\times 10^{-2}$\\
0.010&$6.49\times 10^{-25}$ &0.45&$4.76\times 10^{-2}$\\
0.011&$6.83\times 10^{-24}$ &0.5&$9.24\times 10^{-2}$\\
0.012&$5.47\times 10^{-23}$ &0.6&$2.74\times 10^{-1}$\\
0.013&$3.52\times 10^{-22}$ &0.7&$6.50\times 10^{-1}$\\
0.014&$1.88\times 10^{-21}$ &0.8&$1.32\times 10^{0}$\\
0.015&$8.64\times 10^{-21}$ &0.9&$2.39\times 10^{0}$\\
0.016&$3.48\times 10^{-20}$ &1&$3.99\times 10^{0}$\\
0.018&$4.08\times 10^{-19}$ &1.25&$1.10\times 10^{1}$\\
0.020&$3.40\times 10^{-18}$ &1.5&$2.39\times 10^{1}$\\
0.025&$2.35\times 10^{-16}$ &1.75&$4.40\times 10^{1}$\\
0.030&$5.93\times 10^{-15}$ &2&$7.27\times 10^{1}$\\
0.040&$6.44\times 10^{-13}$ &2.5&$1.59\times 10^{2}$\\
0.050&$1.79\times 10^{-11}$ &3&$2.89\times 10^{2}$\\
0.060&$2.23\times 10^{-10}$ &3.5&$4.64\times 10^{2}$\\
0.070&$1.67\times 10^{-9}$ &4&$6.84\times 10^{2}$\\
0.080&$8.73\times 10^{-9}$ &5&$1.26\times 10^{3}$\\
0.090&$3.52\times 10^{-8}$ &6&$2.00\times 10^{3}$\\
0.10&$1.17\times 10^{-7}$  &7&$2.88\times 10^{3}$\\
0.11&$3.33\times 10^{-7}$  &8&$3.90\times 10^{3}$\\
0.12&$8.38\times 10^{-7}$  &9&$5.03\times 10^{3}$\\
0.13&$1.91\times 10^{-6}$  &10&$6.25\times 10^{3}$\\
\hline
\end{tabular}\label{tab2}
\end{table}

\begin{figure}[htbp]
\includegraphics[width=10 cm]{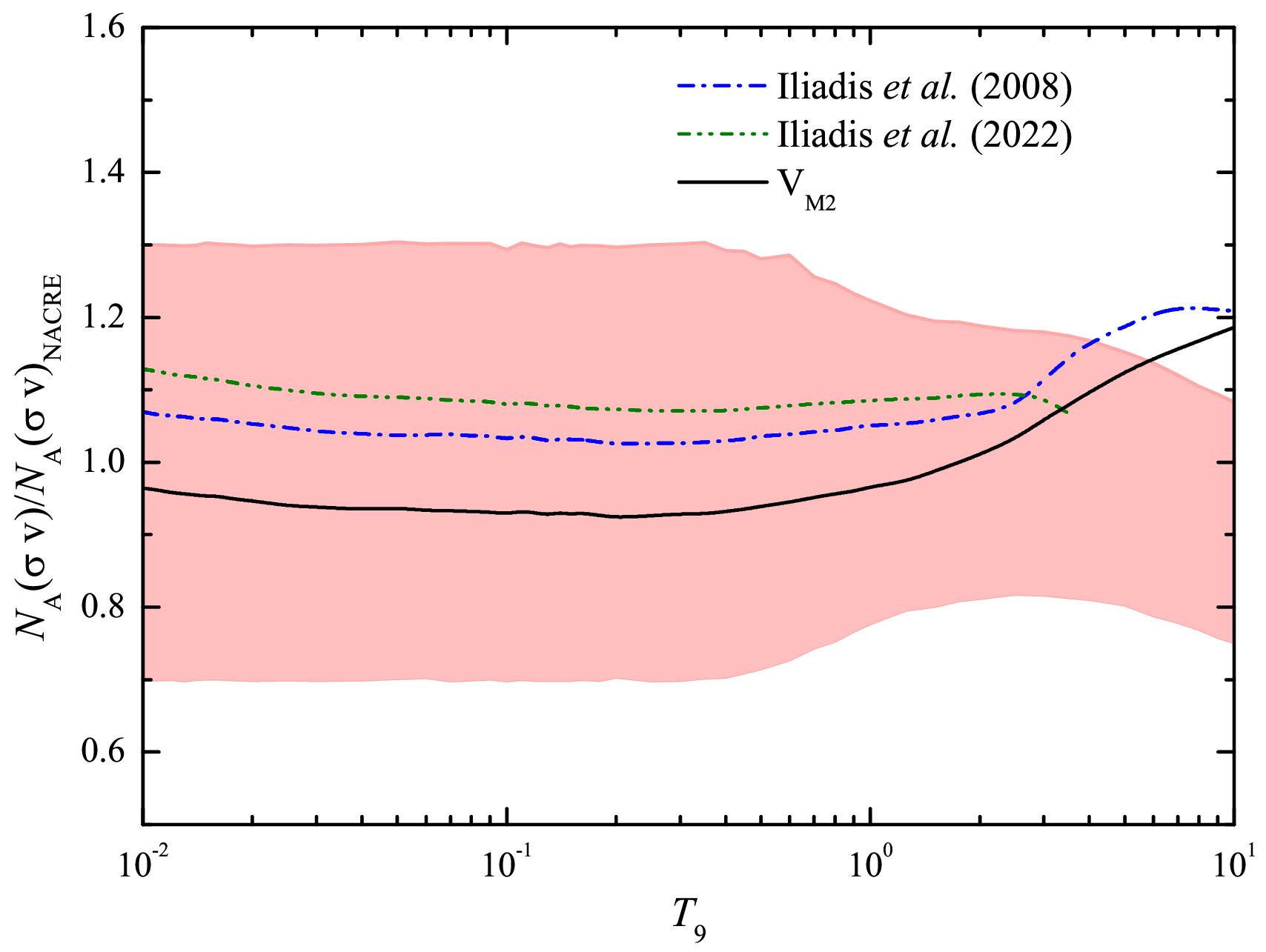}\caption{The $^{16}$O(p,$\gamma)^{17}$F
astrophysical reaction rates estimated within the potential model
$\textrm{V}_\textrm{M2}$ in comparison with the results of
Refs.~\cite{il08,il22}, normalized to the NACRE rate
\cite{NACRE99}.} \label{fig7}
\end{figure}

The numerical values of the estimated reaction rates within the
potential model $\textrm{V}_\textrm{M2}$  for the
$^{16}$O(p,$\gamma)^{17}$F direct capture process are given in Table
~\ref{tab2}. In Fig.~\ref{fig7} the calculated reaction rates
normalized to the NACRE rate \cite{NACRE99}, are compared to the
results of Refs. \cite{il08,il22} in the temperature range from
$T_9$=0.001 to $T_9$=10, where $T_9$ is the temperature in the
10$^9$ K. As can be seen from the figure, the new reaction rates
obtained in the present work are in very good agreement with
available data obtained within the phenomenologic R-matrix approach
\cite{il08} up to the $T_9$=10 and the Bayesian approximation
\cite{il22} up to the $T_9$=1 temperature region. The potential
model reproduces the both absolute values and temperature dependence
of the reaction rates of these models.

\section{Conclusion}

The astrophysical direct nuclear capture process
$^{16}$O(p,$\gamma)^{17}$F was studied within the potential model
approach. The Woods-Saxon form of the phase-equivalent potentials
for the p$-^{16}$O interaction has been examined in the description
of the experimental data for the astrophysical $S$ factor and the
reaction rates of the process. The potential models reproduce the
binding energies and the empirical values of ANC from different
sources for the $^{17}$F(5/2$^+$) ground and $^{17}$F(1/2$^+$)
($E^*$=0.495 MeV) excited bound states. The best description of the
experimental data for the astrophysical $S$ factor is obtained
within the potential model $\textrm{V}_\textrm{M${\rm{2}}$}$ which
yields the ANC values 1.043 fm$^{-1/2}$ and 75.484 fm$^{-1/2}$
\cite{artem09} for the $^{17}$F($5/2^{+}$) ground and
$^{17}$F($1/2^{+}$) excited bound states, respectively. It was
demonstrated that a dominant contribution to the astrophysical $S$
factor of the direct capture process is produced by the $E$1
transition from the initial $P$-waves. The zero-energy astrophysical
factor $S(0)=9.321$ KeV b was estimated by using the asymptotic
expansion method. The calculated reaction rates are in good
agreement with the R-matrix approach and the Bayesian model within
the temperature region up to 10$^{10}$ K.

\section*{Acknowledgements}
The authors thank A.S. Kadyrov for useful discussion of the
presented results.

 \end{document}